\documentclass[12pt]{spieman}  
\usepackage{amsmath,amsfonts,amssymb}
\usepackage{graphicx}
\usepackage{setspace}
\usepackage{tocloft}

\usepackage{caption} 

\usepackage{lineno}

\title{Two-step focusing anamorphic catadioptric telescope and componentwise aberration compensations}

\author[]{Dmitry Zhuridov
	\footnote{E-mail address: dmitry.zhuridov@uwr.edu.pl}}
\affil[]{University of Wroc\l aw, Institute of Theoretical Physics, Faculty of Physics and Astronomy, Plac Maxa Borna 9, Wroc\l aw, Poland, 50204}

\cftpagenumbersoff{figure}
\cftpagenumbersoff{table} 
\begin{document} 
	\maketitle

\begin{abstract}
Aberration compensation with emphasis on the generalized spherical aberration components is discussed for plane-symmetric and anamorphic optical systems. A narrow field-of-view double-plane symmetric telescope objective containing reflective and refractive subsystems, which focus light separately in the two dimensions of the image plane, is introduced. Analytical derivation of the compensation rules is discussed for the telescope being composed of a Cassegrain-type cylindrical reflective subsystem and an economical refractive one that is determined by up to six essential parameters. Performance of a more advanced version of this telescope, which contains a less aberrative nine-parameter refractive subsystem, is studied using the ray-tracing simulator CODE~V. The telescopes of the introduced type provide two independent focal distances for the discussed focusings in the orthogonal dimensions allowing to set up a favorable field of view, and potentially can be used to achieve larger effective aperture area and resolution on a small spacecraft.
\end{abstract}

\keywords{anamorphic optical system, spherical aberration, catadioptric system, Cassegrain telescope, cylindrical mirror, toroidal lens}


\begin{spacing}{2}   

\section{Introduction}\label{sect:intro}  

Anamorphic optical systems (OS), which possess a double-plane symmetry, have been investigated for many years~\cite{Chretien,A_prism_1978,Jung_Lee_1997,Neil_2006,Peschel:2014,Kashima:2017yub,Shi_Zheng} mainly due to their rich capabilities for image formation. However, they reveal also copious structure of  aberrations~\cite{Chretien,Wynne,Slusarev,Rusinov,Neil_2006,Yuan-Sasian_I,Yuan-Sasian_II,Caron-Baumer,zhuridov1} that remain a subject of research. In particular, S.~Yuan and J.~Sasian have argued that, in the paraxial region, an anamorphic system can be treated as two associated rotationally symmetric OS, and derived the generalized ray-tracing equations that include the aberration terms up to the third order~\cite{Yuan-Sasian_I,Yuan-Sasian_II}. Recently, set of the third-order aberration coefficients in the paraxial approximation was obtained for the systems of cylindrical mirrors or/and lenses directly from the calculation of the optical path difference without approximating these OS by the pairs of rotationally symmetric OS, and as a particular example a cylindrical analog of the Cassegrain telescope was considered~\cite{zhuridov1}. The present research is devoted to its possible anamorphic extensions. 
Historically, anamorphic systems have been used in wide-screen imaging since 1950s~\cite{Neil_2006}, 
in laser beam profile controlling since 1970s~\cite{A_prism_1978}, 
in semiconductor chip inspection since 1990s~\cite{Jung_Lee_1997}, etc. 
However, in this article we are interested in yet another possible type of anamorphic solutions, which allow for two large magnifications in the orthogonal directions at the image plane and good control of the field of view. 
These solutions can help in fitting a 
telescope, which collecting area is larger in one dimension, into a space restricted payload or fairing. Rotating such an anamorphic telescope, which provides high spatial resolution in 
1D, one can obtain high resolution 2D images after post processing~\cite{Zhuridov:patent}. 
This gives effectively a larger collection area and resolution as compared to what is achievable with a rotationally symmetric design.

We start from a brief discussion of third-order aberrations for either anamorphic or plane-symmetric OS, concentrating on the generalized spherical aberration, in Sect.~\ref{sec:2}. 
Next, we introduce the concept of a catadioptric telescope with two independent focal distances in Sect.~\ref{sec:3}. We impose spherical aberration compensation in the XZ and YZ planes in its initial design and study performance using a ray-tracing software. We discuss our findings and conclude in Sect.~\ref{sec:4}.

\section{Plane symmetries and aberrations}\label{sec:2}

The transverse primary aberrations for paraxial rays passing in the Z direction through a generic system of cylindrical optical surfaces, which is symmetric under reflection in the YZ plane and translationally invariant along the Y axis, were obtained in Ref.~\cite{zhuridov1} by 
applying a classical method based on direct calculation of the optical path difference~\cite{Schroeder}. We consider generalization of this OS to a single-plane or double-plane symmetric one. In the later case the Z axis coincides with the crossing line of the two symmetry planes, and the OS may contain anamorphic surfaces, e.g., toroidal ones. The wave aberration function $W$ being the optical path difference between real and intended ideal wavefronts can be expressed in terms of the four transverse coordinates \{$x_0$, $y_0$, X, Y\} of the points $P_0(x_0,y_0,0)$ and $P(x,y,z)$, in which arbitrary ray crosses the object plane and the exit pupil plane, respectively~\cite{Born_Wolf}. The both planes are orthogonal to the Z axis. 
Symmetry in either XZ or YZ plane as well as double-plane symmetry in both these planes leaves the following six second order invariants for the considered four variables
\begin{eqnarray}
	x_0^2, \quad y_0^2, \quad x^{2}, \quad y^{2}, \quad x_0x, \quad y_0y.
\end{eqnarray}
The fourth-order wave aberration function is a generic quadratic form in these invariants. The number of its terms can be calculated using the relation between the 2-combinations with repetitions and the binomial coefficient as follows
\begin{flalign}
	\left(
	\begin{pmatrix}
		6\\
		2
	\end{pmatrix}\right) 
	= \begin{pmatrix}
		6+2-1\\
		2
	\end{pmatrix} = \frac{7\times6}{2} = 21.
\end{flalign}
This form contains three constant piston terms (involving only $x_0$ and $y_0$) and two pairs of identical terms with either $x_0^2x^2$ or $y_0^2y^2$. 
This leaves in the fourth-order wave aberration function~\cite{burfoot,barakat_houston,yuan_PhD}
\begin{flalign}\label{eq:W4}
	W^{(4)}\! &= a_1x^{4} + a_2y^{4} + a_3x^2y^{2} \nonumber\\
	&+ a_4x_0x^{3} + a_5y_0x^{2}y + a_6x_0x y^{2} + a_7y_0y^{3} \nonumber\\
	&+ a_8x_0^2x^{2} + a_9y_0^2y^{2} + a_{10}y_0^2x^{2} + a_{11}x_0^{2}y^{2} + a_{12}x_0y_0x y \nonumber\\
	&+ a_{13}x_0^3x + a_{14}y_0^3y + a_{15}x_0y_0^2x + a_{16}x_0^2y_0y
\end{flalign}
sixteen different terms that generate third-order aberrations of spherical, comatic, astigmatic (plus field curvature) and distortion types, corresponding to the four lines in~\eqref{eq:W4}. For paraxial rays the components of the third-order transverse aberrations can be calculated as
\begin{eqnarray}
	\delta_{1x}
	= - \frac{D}{n}\frac{\partial W^{(4)}}{\partial x},  \qquad  
	\delta_{1y}
	= - \frac{D}{n}\frac{\partial W^{(4)}}{\partial y},
\end{eqnarray}
where $D$ is the distance between the exit pupil plane and the image plane, and $n$ is the refractive index of the material behind the exit pupil. 
In particular, the generalized spherical (GS) 
aberrations are given by 
\begin{eqnarray}
	n\delta_{1x}^{GS} &=&  (\tilde a_1x^{2} + \tilde a_3y^{2})x, \label{eq:delta_x}\\
	n\delta_{1y}^{GS} &=&  (\tilde a_3x^{2} + \tilde a_2y^{2})y, \label{eq:delta_y}
\end{eqnarray}
where $\tilde a_{1,2}=4a_{1,2}D$, $\tilde a_3=2a_3D$. Clearly, in the case of a single-plane or double-plane symmetry GS aberrations depend on three parameters, which is similar to the case of a single-plane symmetry combined with translational invariance~\cite{zhuridov1}. The GS aberration degenerates into ordinary spherical aberration for $\tilde a_1=\tilde a_2 = \tilde a_3$. Hence, for the rays in the XZ or YZ planes the coefficients $\tilde a_1$ and $\tilde a_2$, respectively, play the role of a usual spherical aberration coefficient.

We notice that in the case of no symmetry constraints the wave aberration function can contain generically five fourth-order power products of the variables X and Y: $x^4$, $x^3y$, $x^2y^2$, $xy^3$ and $y^4$. Therefore five parameters can enter GS aberrations.

\begin{figure}
	\centering\includegraphics{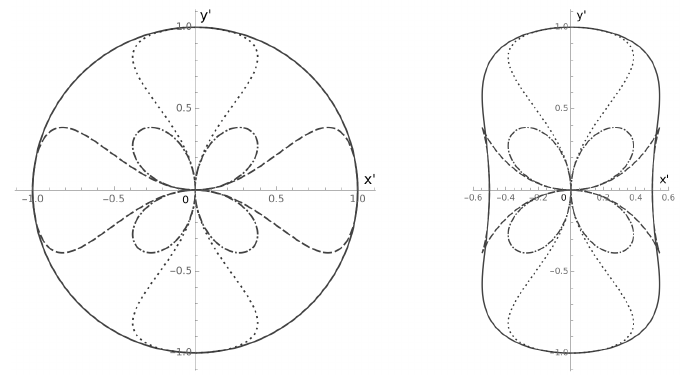}
	\caption
	{Generalized spherical aberrations for $a=1$ and the following chosen values for the parameter set \{$\tilde a_{1}\rho^3$, $\tilde a_{2}\rho^3$\} in Eqs.~(\ref{eq:delta_x})-(\ref{eq:polar}). {\it Left}:  \mbox{\{1, 1\}} ({\it solid}), \{1, 0\} ({\it dashed}), \{0, 1\} ({\it dotted}) and \{0, 0\} ({\it dot-dashed}). {\it Right}:  \{$\tilde a_{1}\rho^3$, $\tilde a_{2}\rho^3$\} set: \mbox{\{0.5, 1\}} ({\it solid}), \mbox{\{0.5, 0\}} ({\it dashed}), \mbox{\{0, 1\}} ({\it dotted}) and \{0, 0\} ({\it dot-dashed}).}
	\label{fig:1}
\end{figure}

In the polar coordinates of the exit pupil,
\begin{eqnarray}\label{eq:polar}
	x=\rho\cos\theta, \qquad y=\rho\sin\theta,
\end{eqnarray}
for the rays in the XZ and YZ planes the angle $\theta$ is fixed and equal to $k\pi$ and $\pi/2+k\pi$, respectively, where $k$ is an integer. Then conditions for GS aberration compensation in these planes nullify the coefficients $\tilde a_1$ and $\tilde a_2$. The remaining coefficient $\tilde a_3$ enters the 
quadrifolium (clover) component of GS aberrations:
\begin{flalign}
	n\delta_{1x}^{GS} &= 
	a \sin^2\theta \cos\theta 
	= \frac{a}{2} \sin 2\theta \sin\theta, \label{eq:ellip_ab1}\\
	n\delta_{1y}^{GS} &=  
	a \cos^2\theta \sin\theta 
	= \frac{a}{2} \sin 2\theta \cos\theta, \label{eq:ellip_ab2}
\end{flalign}
where $a=\tilde a_3\rho^{3}$, as shown in Fig.~\ref{fig:1} that is drawn with Mathematica~\cite{Mathematica}. 

\section{Generalized spherical aberration for anamorphic catadioptric telescope}\label{sec:3}

In this section we present an anamorphic telescope~\cite{Zhuridov:patent} with a separate focusing along the orthogonal directions X and Y, and the optical axis coinciding with the Z axis. To be more specific, we discuss catadioptric schemes consisting of the two sets of optical elements: reflective set (I) and refractive set (II). The set (I) contains cylindrical mirror(s) and is responsible for focusing collimated light, which is parallel to the Z axis, to a focal segment (in case of absence of the second set) at the focal distance $F_1$. The set (II) with smaller focal distance $F_2$ contains anamorphic lenses and is responsible for compressing the light pencil in the direction of the focal segment and its focusing to a focal point. The set (II), being an anamorphic extension for the set (I), breaks the translational invariance of the OS to the second plane symmetry. 

\begin{figure}[ht!]
	\centering\includegraphics{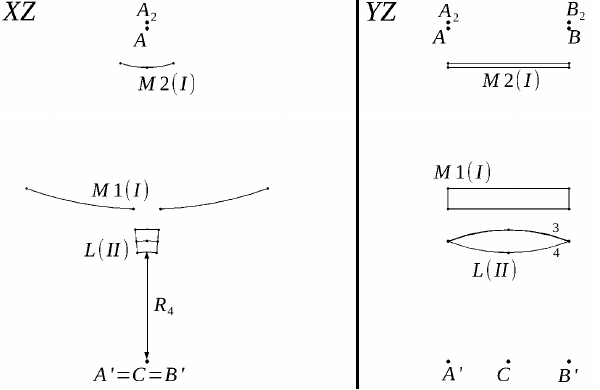}

	\vspace{3mm}
	\caption
	{The discussed anamorphic OS with a single-lens set (II) in the projections of XZ ({\it left}) and YZ~({\it right}).}
	\label{fig:OS}
\end{figure}

For definiteness we take the translationally invariant set (I) of Cassegrain type~\cite{zhuridov1}, which consists of the two cylindrical mirrors: the concave primary M1 and the convex secondary M2, as shown in Fig.~\ref{fig:OS}. The focal segment $AB$ of the primary M1 is parallel to the focal segment $A_2B_2$ of the secondary M2. 
In our analytical consideration in the next two subsections~\ref{sec:singlet} and \ref{sec:doublet} we concentrate on the economical examples of the set (II) containing either two or three optical surfaces, respectively. Then we perform ray-tracing for a more advanced (and less aberrative) version of the discussed telescope with a four-surface set (II) in the subsection~\ref{sec:proof}, and discuss its field of view in the subsection~\ref{sec:FOV}.
%

\subsection{Single lens set (II)}\label{sec:singlet}

\begin{figure}[ht!]
	\centering\includegraphics{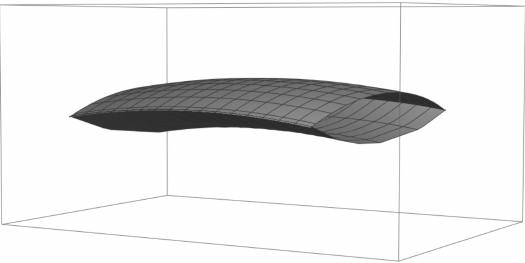}

	\vspace{2mm}
	\caption
	{A schematic isometric view of the optical surfaces 3 (upper) and 4 (lower) of the toroidal lens L. The lens thickness is discussed in section~\ref{sec:YZplane}, its edges are not sharp in practice.}
	\label{fig:lens}
\end{figure}

We add a generalized toroidal shape lens L behind the mirror set (I) in such a way that the symmetry axis of the respective torus coincides with the focal line $A^\prime B^\prime$ of the 
set (I). An example lens L is shown in Fig.~\ref{fig:lens} for the case of minimal thickness (sharp edges) that is not necessary. This lens is a body of rotation of its YZ profile with respect to the axis $A^\prime B^\prime$. We choose L with a biconvex YZ profile. The OS focal point $C$ is in the center of the segment $A^\prime B^\prime$. The entire OS contains only four optical surfaces, indexed by 1 and 2 for M1 and M2, respectively, and 3 and 4 for the surfaces of L. We describe the cylindrical surfaces by the equation
\begin{eqnarray}\label{eq:surface}
	z =  \pm
	\left[\frac{x^2}{2R_i} + (1+\kappa_i)\frac{x^4}{8R_i^3}\right],
\end{eqnarray}
where overall sign determines the direction of surface convexity, $R_i$ is the $i$-th surface's radius of curvature (always positive) and its only asphericity is in conic constant $\kappa_i$.

In the following we discuss aberrations for the considered anamorphic OS, concentrating on GS aberration for brevity. We introduce the conditions for compensation of its components in the XZ and YZ planes, which correspond to the dotted and dashed lines in Fig.~\ref{fig:1}, respectively, where the dot-dashed line corresponds to the fulfillment of the both conditions at once. We remark that the introduced method can be applied also to reduction of comatic and other aberration types.

\subsubsection{XZ plane}
Following the method of Ref.~\cite{zhuridov1}, we determine contribution to GS aberration for the considered system for paraxial rays in the XZ plane and the marginal ray height at the primary (indicated by the superior `(1)') by means of the following function
\begin{flalign}
	A_{3s}^{(1)} = A_{311} + A_{312} \left( \frac{x_2}{x_1} \right)^4\!\! + A_{313} \left( \frac{x_3}{x_1} \right)^4\!\!  + A_{314} \left( \frac{x_4}{x_1} \right)^4\!,
\end{flalign}
where $x_i$ is the marginal ray height at the $i$-th surface, the first subindex of the `$A$' quantities refers to spherical aberration ($j=3$ in terms of Refs.~\cite{zhuridov1,Schroeder}), the subindex $s$ goes for (whole optical) system, the second subindex of the aberration coefficients $A_{ijk}$ indicates the aberration component related to the XZ plane, and the third subindex of $A_{ijk}$ numbers the optical surface. The aberration coefficients for the mirrors are given by
\begin{flalign}
	A_{311} = -\frac{\kappa_1+1}{4R_1^3}, \quad  A_{312} = \frac{1}{4R_2^3} \left[ \kappa_2 + \left( \frac{\mu+1}{\mu-1} \right)^2 \right],
\end{flalign}
where $\mu>0$ is the transverse magnification of the 
secondary.

To derive the aberration coefficients $A_{313}$ and $A_{314}$ we replace the refractive surfaces 3 and 4 by the cylindrical ones, 
which are tangential to them in the XZ plane and are described by~\eqref{eq:surface} with the overall plus sign. Then we use the following generic expression~\cite{zhuridov1}
\begin{flalign}
	A_{31i} = -\, \frac{1}{8} \!\left[ \frac{\kappa}{R^3}(n_1-n_0) + 
	\frac{n_0}{s_0} \left(1+\frac{n_0}{n_1}\frac{s_0}{s_1}\right)\! \left( \frac{1}{s_{0}} + \frac{1}{R}\right)^2\right]_i,
	\label{eq:A31}
\end{flalign}
where the $i$ index outside the quadratic parenthesis means that all the quantities inside that parenthesis refer to the $i$-th surface. 
The XZ projections for the 3rd and 4th surfaces are the circle arcs with the radii $R_3$ and $R_4$, respectively, which are centered in the OS focal point $C$, see Fig.~\ref{fig:OS} (left). We take the refractive index to be $n$ inside the lens and unity outside. For $\kappa_{3x}=\kappa_{4x}=0$, where the index X refers to the section by the XZ plane, 
we obtain
\begin{eqnarray}
	A_{313} &=& -\, \frac{1}{8s_{03}} \left(1+\frac{s_{03}}{ns_{13}}\right) \left( \frac{1}{s_{03}} + \frac{1}{R_3}\right)^2, \label{eq:A313}\\
	A_{314} &=& -\, \frac{n}{8s_{04}} \left(1+\frac{ns_{04}}{s_{14}}\right) \left( \frac{1}{s_{04}} + \frac{1}{R_4}\right)^2,\label{eq:A314}
\end{eqnarray}
where $s_{0i}$ and $s_{1i}$ are the lengths of the XZ projections of the 
chief ray segments connecting the $i$-th surface with the respective object and image point. 
Using the rule of `negative distance' between a surface and a virtual object (converging incoming beam), we substitute 
\begin{eqnarray}
	s_{0i}=-s_{1i}=-R_i	\label{eq:s0s1}
\end{eqnarray} 
with $i=3,4$ in Eqs.~(\ref{eq:A313}) and (\ref{eq:A314}), so the coefficients $A_{313}$ and $A_{314}$ vanish. Then condition of compensation of GS aberration in the XZ plane coincides with the one for isolated set (I)~\cite{zhuridov1}:
\begin{flalign}
	F_{XZ} 
	\equiv \kappa_1 + 1 - \frac{k_x^4}{\rho_{21}^3} \left[ \kappa_2 + \left( \frac{\mu+1}{\mu-1} \right)^2 \right] = 0,
\end{flalign}
where $\rho_{21} = R_2/R_1$ and $k_x = x_2/x_1$. In particular, for a telescope of classical Cassegrain type: $\kappa_1=-1$ (
parabolic M1 profile) and $\kappa_2=-(\mu+1)^2(\mu-1)^{-2}<-1$ (
hyperbolic M2 profile).

\subsubsection{YZ plane}\label{sec:YZplane}
The GS aberration component in the YZ plane for the marginal ray height at the primary can be determined by
\begin{flalign}\label{eq:B3s}
	B_{3s}^{(1)} = \tilde B_{31} + \tilde B_{32} \left( \frac{y_2}{y_1} \right)^4\!\! + B_{33}^* \left( \frac{y_3}{y_1} \right)^4\!\!  + B_{34}^* \left( \frac{y_4}{y_1} \right)^4\!,
\end{flalign}
where $y_i$ is the marginal ray distance from the Z axis at the $i$-th surface, the first subindex of all $B$s goes for spherical aberration, while their second subscript either refers to the system `$s$' or numbers the optical surface. The aberration coefficients $\tilde B_{3i}\equiv A_{32i}$ vanish. (The coefficients referred to cylindrical surfaces, which YZ profiles are linear, we mark with tildes as in Ref.~~\cite{zhuridov1}, and the coefficients referred to generalized toroidal surfaces we mark with stars.) 
To obtain the coefficients $B_{33}^*$ and $B_{34}^*$ we use the cylindrical surfaces, which are tangential to the surfaces 3 and 4 in the YZ plane and translationally invariant along the X axis, in contrast to the translation along the Y axis in Ref.~\cite{zhuridov1}. To indicate this difference we replace squiggles with stars. 
This follows in the interchanged roles of $A\leftrightarrow B$ and $x_i\leftrightarrow y_i$, 
which results in $B_{3i}^*=A_{31i}$~\cite{zhuridov1}. 
By labeling curvature radii in the YZ plane with lowercase letters, according to the proper overall sign in~\eqref{eq:surface}, we obtain
\begin{flalign}
	B_{33}^* = -\frac{\kappa_{3}(n-1)}{8r_3^3} - 
	\frac{1}{8s_{03y}} \left(1+\frac{s_{03y}}{ns_{13y}}\right)\! \left( \frac{1}{s_{03y}} + \frac{1}{r_3}\right)^2\!, \label{eq:B33}\\
	B_{34}^* = -\frac{\kappa_{4}(n-1)}{8r_4^3} - 
	\frac{n}{8s_{04y}} \left(1+\frac{ns_{04y}}{s_{14y}}\right)\! \left( \frac{1}{s_{04y}} - \frac{1}{r_4}\right)^2\!, \label{eq:B34}
\end{flalign}
where $r_i\equiv r_{iy}$ and $\kappa_i\equiv \kappa_{iy}$ refer to the surface that is tangential to the $i$-th surface, and the quantities relevant to the YZ plane are indexed by Y. 

Paraxial equations for refraction on the discussed surfaces tangential to 3 and 4 read
\begin{eqnarray}\label{eq:lens}
	\frac{n}{s_{13y}} - \frac{1}{s_{03y}} = \frac{n-1}{r_3},  \qquad  \frac{1}{s_{14y}} - \frac{n}{s_{04y}} = \frac{n-1}{r_4},
\end{eqnarray}
where we substitute $s_{03y}=\infty$ and $s_{04y}=s_{13y}-d_L$ ($\kappa_i=0$ in the paraxial approximation). The central thickness $d_L$ of 
the lens L 
is given by 
\begin{eqnarray}
	d_L = R_3-R_4,
\end{eqnarray}
where the capital `$R$' denote the radii of curvature in the XZ plane, according to our notation. 
It can not be smaller than the sum of sags $d_3$ and $d_4$, which can be expressed as
\begin{eqnarray}\label{eq:di}
	d_i(r_i,L_y) = r_i \left[ 1 - \sqrt{1-\left(\frac{L_y}{2r_i}\right)^2} \right], \ \ i=3,4,
\end{eqnarray}
where $L_y=AB$ is the lens size along the Y axis (in Fig.~\ref{fig:OS} $d_L=d_3+d_4$). Using Eq.~(\ref{eq:lens}), 
Eqs.~(\ref{eq:B33}) and (\ref{eq:B34}) can be rewritten as
\begin{flalign}
	B_{33}^* &= -\, \frac{n-1}{8r_3^3} \left(\kappa_{3}+\frac{1}{n^2}\right), \label{eq:B33new}\\
	B_{34}^* &= -\, \frac{n-1}{8r_4^3} \left[\kappa_{4} + \frac{n}{\nu^3} \left(1 + n^2 + n\nu\right)  \right] \left( 1-n + \nu \right)^2  \label{eq:B34new}
\end{flalign}
with the new dimensionless function
\begin{flalign}\label{eq:nu}
	\nu = n\,\frac{r_3}{r_4} - (n-1)\,\frac{d_L}{r_4}.
\end{flalign} 

\begin{figure}[ht!]
	\centering\includegraphics{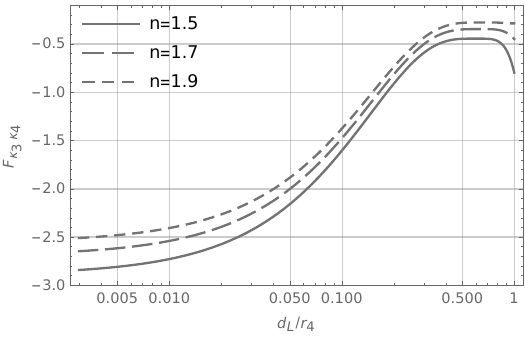}
	\caption
	{The function of conic constants $F_{\kappa_3\kappa_4}$ versus the ratio of the lens thickness $d_L$ to the radius of curvature $r_4$ for $r_3=r_4$ and the chosen values of the refractive index $n$: 1.5, 1.7 and 1.9.}
	\label{fig:Fk3k4}
\end{figure}

\begin{figure}[ht!]
	\centering\includegraphics{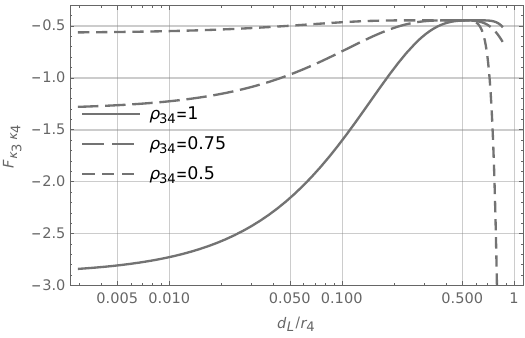}
	\caption
	{The function of conic constants $F_{\kappa_3\kappa_4}$ versus the ratio of the lens thickness $d_L$ to the radius of curvature $r_4$ for the refractive index $n=1.5$ and the chosen values of the ratio of radii $\rho_{34}$: 1, 0.75 and 0.5.}
	\label{fig:Fk3k4_n15}
\end{figure}

\begin{figure}[ht!]
	\centering\includegraphics{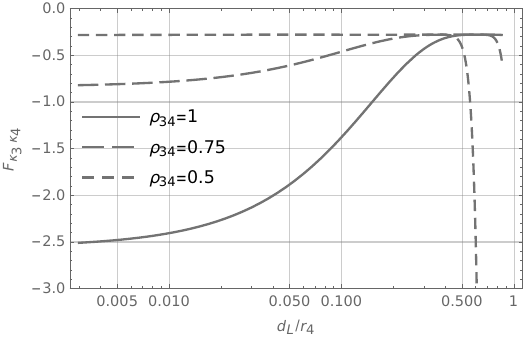}
	\caption
	{The function of conic constants $F_{\kappa_3\kappa_4}$ versus the ratio of the lens thickness $d_L$ to the radius of curvature $r_4$ for the refractive index $n=1.9$ and the chosen values of the ratio of radii $\rho_{34}$: 1, 0.75 and 0.5.}
	\label{fig:Fk3k4_n19}
\end{figure}

According to Eqs.~(\ref{eq:B3s}), (\ref{eq:B33new}) and (\ref{eq:B34new}), this is convenient to describe the discussed aberration cancellation in terms of the function
\begin{flalign}
	F_{\kappa_3\kappa_4} = \kappa_3 + \kappa_4 \left(\frac{r_3}{r_4}\right)^3 \left(\frac{y_4}{y_3}\right)^4,
\end{flalign}
which is linear in terms of the conic constants and vanishes along with them, i.e. for spherical lens. 
Denoting $\rho_{34} \equiv r_3/r_4$ and taking into account that
\begin{flalign}
	\frac{y_4}{y_3} = 1 - \frac{n+1}{n}\,\frac{d_L}{r_3},
\end{flalign}
we rewrite the condition for GS aberration compensation in the YZ plane as
\begin{flalign}
	F_{\kappa_3\kappa_4} &= - \frac{1}{n^2} - \frac{n\rho_{34}^3}{\nu^3} \left(1 + n^2 + n\nu\right) \left( 1-n + \nu \right)^2 \nonumber\\
	&\times \left(1-\frac{n+1}{n\rho_{34}}\frac{d_L}{r_4}\right)^4.
\end{flalign}
In Fig.~\ref{fig:Fk3k4} we show the dependence of this 
function on the ratio $d_L/r_4$ for the case of equal radii $r_3=r_4$ and several choices of $n$. Then in Figs.~\ref{fig:Fk3k4_n15} and \ref{fig:Fk3k4_n19} we plot this dependence for the chosen values of the ratio $\rho_{34}$ and the refractive index of 1.5 and 1.9, respectively. 

Clearly, the function $F_{\kappa_3\kappa_4}$ does not vanish, and therefore both $\kappa_3$ and $\kappa_4$ cannot be zero simultaneously. 
Hence, at least one of the lens surfaces 
must be `aspheric' in its YZ profile in order to make it free of the GS aberration component. This result was expected since the mirrors 1 and 2 do not aberrate the rays in the YZ plane, while the considered toroidal lens effects these rays in the same way as a spherically symmetric one does.


\subsection{Doublet lens set (II)}\label{sec:doublet}

Since two refractive surfaces with vanishing conic constants is not enough to make the set (II) free of GS aberration, we consider its version containing three optical surfaces. Specifically, we choose an anamorphic extension of a refractive doublet objective, contact type, in its standard form with positive front lens and negative rare one, i.e. we take two connected toroidal lenses, positive and negative, with circular YZ profiles. We denote the refractive indices of the positive and negative components by $n$  and $\tilde n$, respectively, and chose $n<\tilde n$. The respective condition for compensation of GS aberrations reads
\begin{flalign}\label{eq:sumB}
	\sum_{i=3}^5 B^*_{3i}\left( \frac{y_i}{y_3} \right)^4=0
\end{flalign}
with
\begin{flalign}
	B_{33}^* &= -\, \frac{n-1}{8n^2r_3^3}, \\
	B_{34}^* &= -\, \frac{n-1}{8r_4^3} \,\frac{n}{\nu^3} \left[ 1 + \frac{n}{\tilde n^2}\left(n - \frac{\tilde n-n}{n-1}\,\nu\right) \right] \left( 1-n + \nu \right)^2, \\
	B_{35}^* &= -\, \frac{\tilde n(\tilde n^2+1)}{8(s_{14}-\tilde d_L)^3},
\end{flalign}
where we took $r_5 = \infty$ (plano-concave negative component) for more concise expression of $B_{35}^*$, $\nu$ is given by Eq.~(\ref{eq:nu}), $d_L$ and $\tilde d_L$ are the thicknesses of positive and negative components of the doublet lens, respectively, and
\begin{flalign}
	s_{14} = \frac{\tilde n \nu r_4}{n(n-1) - (\tilde n-n)\nu}.
\end{flalign}
The distance to the focus behind the doublet lens is given by
\begin{flalign}
	s_{15} = \frac{s_{14}-\tilde d_L}{\tilde n}.
\end{flalign}
The Eq.~(\ref{eq:sumB}) nearly vanishes for a broad range of values for the six parameters, which are essential in this consideration: $r_3$, $r_4$, $d_L$, $\tilde d_L$, $n$ and $\tilde n$.

We remark that the used method can be applied to more complex refractive sets and extended to compensate for other aberrations besides the spherical one.

\begin{figure}[ht!]
	\centering\includegraphics{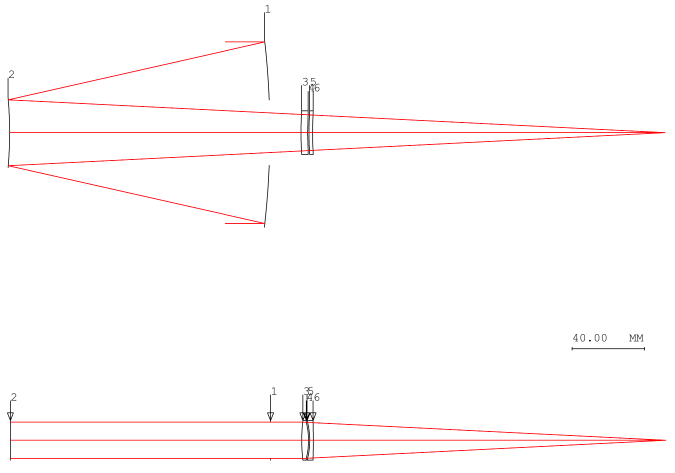}

	\vspace{5mm}
	\caption
	{XZ view (top) and YZ view (bottom) of the considered realization of an anamorphic catadioptric telescope with the RC-type Cassegrain set (I) and the apochromatic doublet set (II).}
	\label{fig:simulation}
\end{figure}
\begin{figure}[ht!]
	\centering\includegraphics{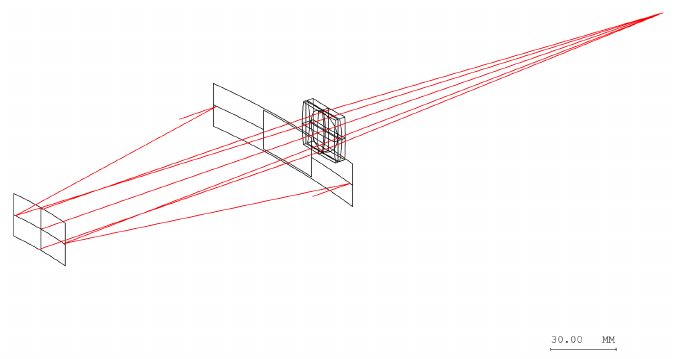}
	\caption
	{Perspective view of the considered realization of an anamorphic catadioptric telescope with the RC-type Cassegrain set (I) and the apochromatic doublet set (II).}
	\label{fig:simulation_per}
\end{figure}

\begin{figure}[ht!]
	\centering\includegraphics{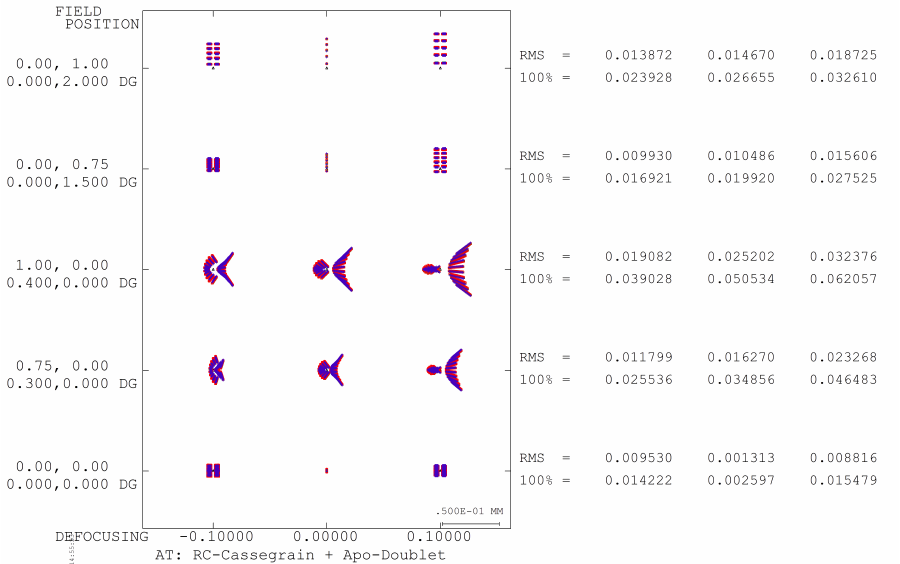} 

	\vspace{5mm}
	\caption
	{Spot diagram for the discussed anamorphic telescope at the image plane for the specified five fields and three focal positions.}
	\label{fig:spot}
\end{figure}
\begin{figure}[ht!]
	\centering\includegraphics{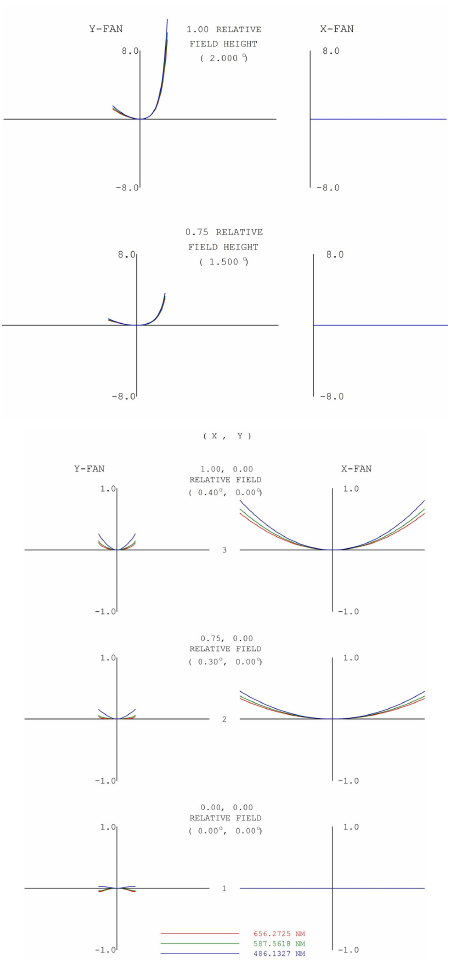}

	\vspace{3mm}
	\caption
	{OPD plot for the discussed anamorphic telescope for the specified five fields.}
	\label{fig:OPD}
\end{figure}

\subsection{Proof of concept by image simulation}\label{sec:proof}

To validate the proposed concept design we consider f/10 anamorphic telescope (AT) with two-mirror set (I) and  doublet lens set (II), which has two equal focal ratios:
\begin{flalign}
	\frac{F_1}{L_x} = \frac{F_2}{L_y} = 10,
\end{flalign}
where $L_x$ and $L_y$ are the sides of its rectangular aperture. To be more specific, we choose the set (I), which is a cylindrical analog of a Ritchey-Chretien type 
Cassegrain. For a set (II) we take toroidal analog of the primary spherical aberration and coma free apochromatic doublet~\cite{telescope-optics}, since the simplified three-surface contact doublet, which is discussed in section~\ref{sec:doublet}, generates essentially larger aberrations. Having the reflective set (I) and apochromatic set (II), we obtain OS that is nearly free of chromatic aberrations. We notice that design optimization for specific narrow observation targets requires a deeper research on selection of a particular set (II).

We simulated this OS and performed ray-tracing using the optical design software CODE~V~\cite{CODE:V}.  The simulation is performed for three wavelengths: d, F, and C (the spectral lines for helium at 587.6~nm and for hydrogen at 486.1 and 656.3 nm, respectively). 
The set (I) parameter values we take as follows: $F_1 = 1000$~mm, $R_1 = 450$~mm, $R_2 = 210$~mm, $\kappa_1=-1.057$ and $\kappa_2=-2.839$. 
Adopting the apochromatic doublet specification~\cite{telescope-optics} to provide the ratio of foci of $F_1/F_2\approx5$, and optimizing consequently the radii of curvature $r_5$ and $r_3$, 
for the set (II) we obtain the following set of curvature radii from $r_3$ to $r_6$ with the sign factors \{$95.75$~mm, $-41.98$~mm, $-41.98$~mm, $-367.74$~mm\} and the lens separation of 0.688~mm, where the positive and negative components' widths are $d_L=3.704$~mm and $\tilde d_L=2$~mm, and their glass types are FPL52\,(OHARA) and ZKN7\,(SCHOTT), respectively. In the above optimization we applied the standard CODE~V procedure with the `Transverse Ray Aberration' error function. In result, performance of the set (II) was improved in both modes of isolation and as part of the AT.

The XZ, YZ and perspective views of the resulting AT are shown in Figs.~\ref{fig:simulation} and \ref{fig:simulation_per}. Its spot diagram is drawn in Fig.~\ref{fig:spot} for three defined focal positions  
and five chosen field points: on-axis $0^\circ$ and two off-axis pairs in the XZ and YZ planes. 
The Y field angles are taken $F_1/F_2$ times larger then the X field angles (this is natural according to section~\ref{sec:FOV}). 
This spot diagram and the following ones are generated by tracing an evenly spaced rectangular 
grid of rays on the entrance pupil from each wavelength with 51 rays across the aperture used for plotting. 
The spot diagrams list, for each field point and zoom position, the following: the relative field heights with X and Y field angles in the left, and the minimum RMS (with the fractional error of $10^{-4}$) and 100\% spot size diameters in the right. 
The corresponding optical path difference (OPD) plot is shown in Fig.~\ref{fig:OPD}. From Figs.~\ref{fig:spot} and \ref{fig:OPD} it is clear that the proposed AT can serve only as a narrow field of view instrument.

\begin{figure}[ht!]
	\centering\includegraphics{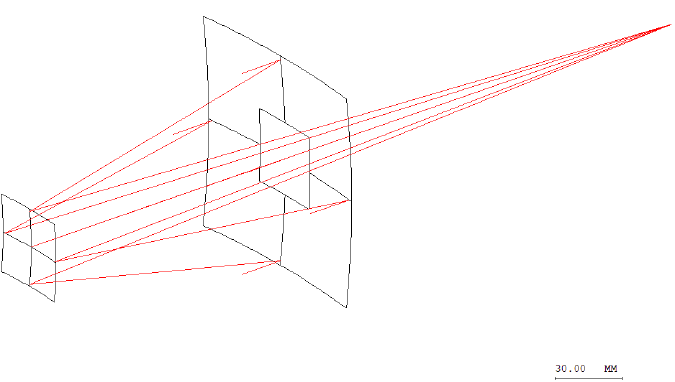}

	\vspace{3mm}
	\caption
	{The perspective view of the axially symmetric analog of isolated set (I) with the aperture 100\,mm\,$\times$\,100\,mm.}
	\label{fig:set1}
\end{figure}
\begin{figure}[ht!]
	\centering\includegraphics{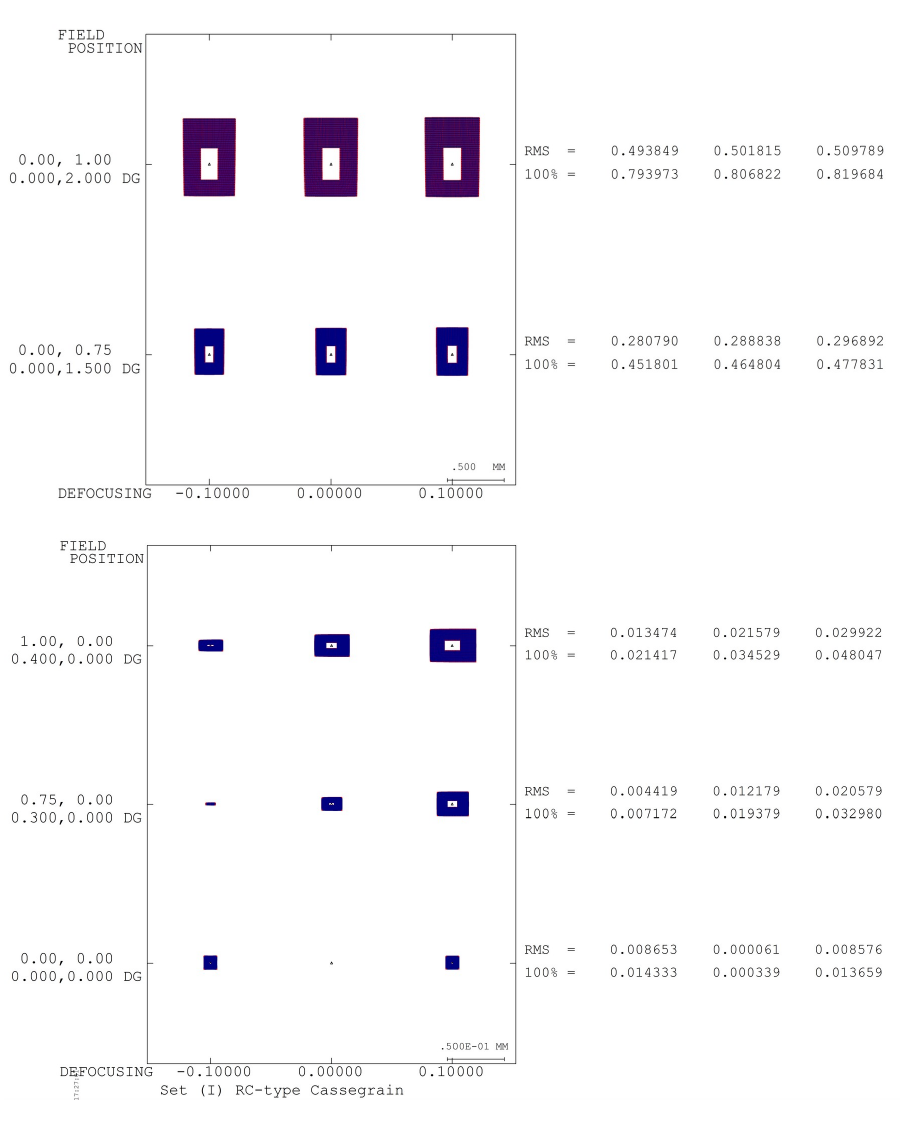}
	\caption
	{The spot diagram for the axially symmetric analog of isolated set (I) with the aperture 100\,mm\,$\times$\,100\,mm.}
	\label{fig:set1spot}
\end{figure}
\begin{figure}[ht!]
	\centering\includegraphics{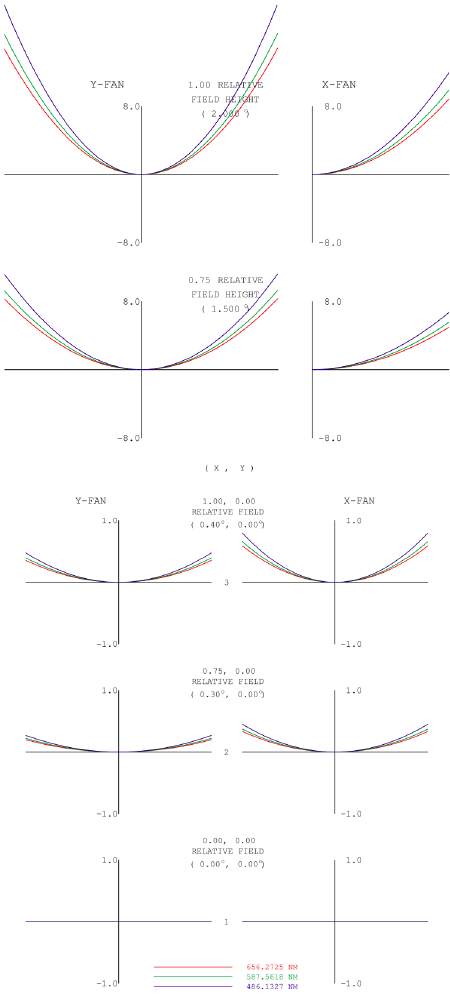}

	\vspace{5mm}
	\caption
	{The OPD plot for the axially symmetric analog of isolated set (I) with the aperture 100\,mm\,$\times$\,100\,mm.}
	\label{fig:set1OPD}
\end{figure}

\begin{figure}[ht!]
	\centering\includegraphics{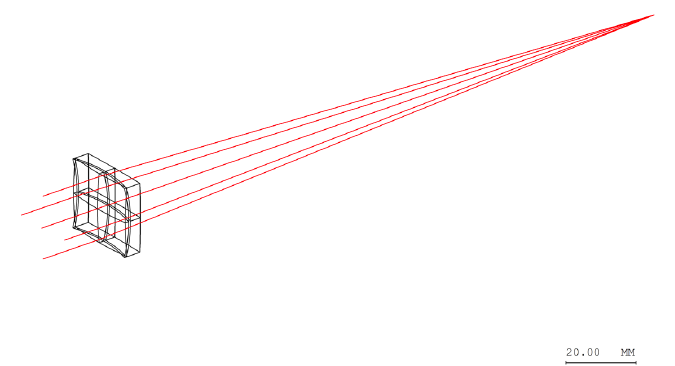}

	\vspace{3mm}
	\caption
	{The perspective view of the axially symmetric analog of isolated set (II) with the aperture 20\,mm\,$\times$\,20\,mm.}
	\label{fig:set2}
\end{figure}
\begin{figure}[ht!]
	\centering\includegraphics{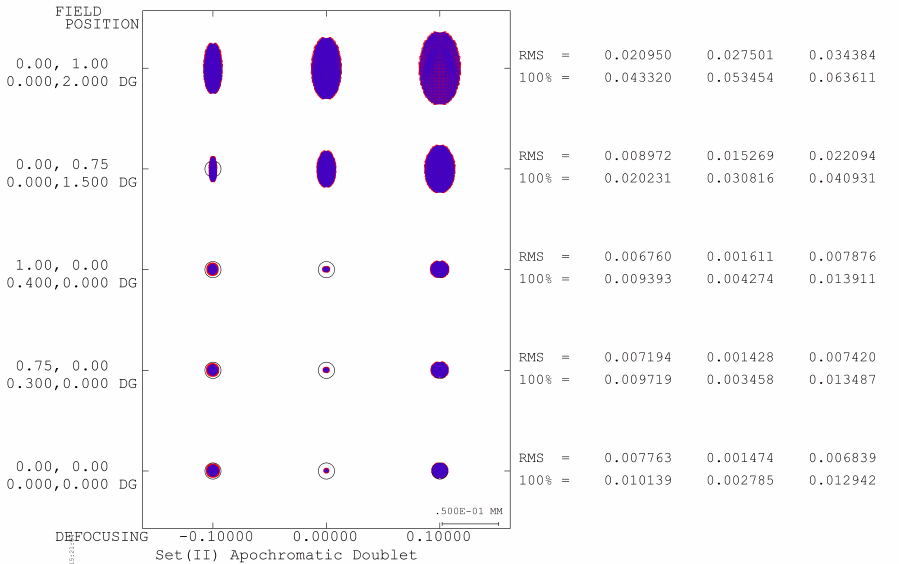}

	\vspace{5mm}
	\caption
	{Spot diagram at the image plane for the axially symmetric analog of isolated set (II) with the aperture 20\,mm\,$\times$\,20\,mm.}
	\label{fig:set2spot}
\end{figure}
\begin{figure}[ht!]
	\centering\includegraphics[width=3in]{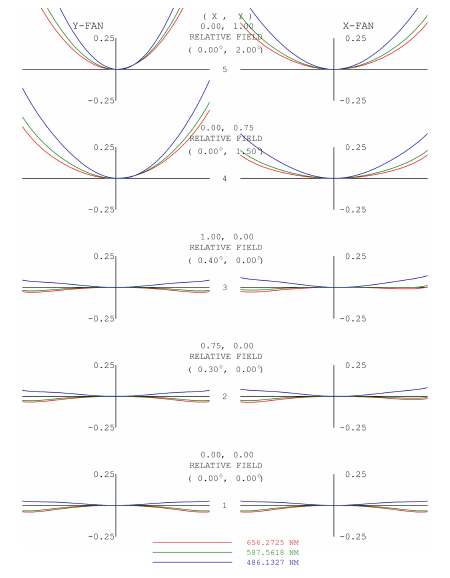}

	\vspace{3mm}
	\caption
	{The OPD plot for the axially symmetric analog of isolated set (II) with the aperture 20\,mm\,$\times$\,20\,mm.}
	\label{fig:set2OPD}
\end{figure}

\begin{figure}[ht!]
	\centering\includegraphics{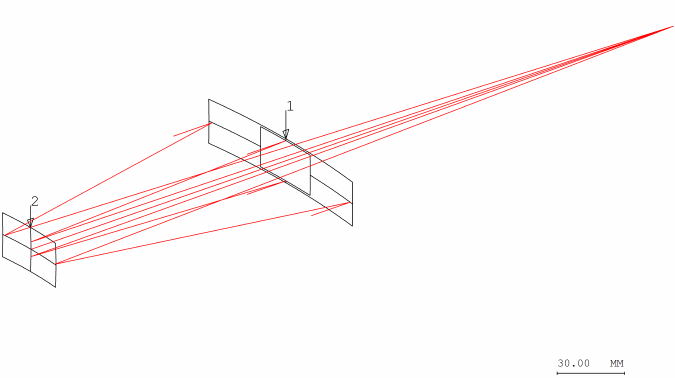}

	\vspace{3mm}
	\caption
	{The perspective view of the axially symmetric analog of isolated set (I) with the rectangular aperture 100\,mm\,$\times$\,20\,mm.}
	\label{fig:set1_rect}
\end{figure}

\begin{figure}[ht!]
	\centering\includegraphics{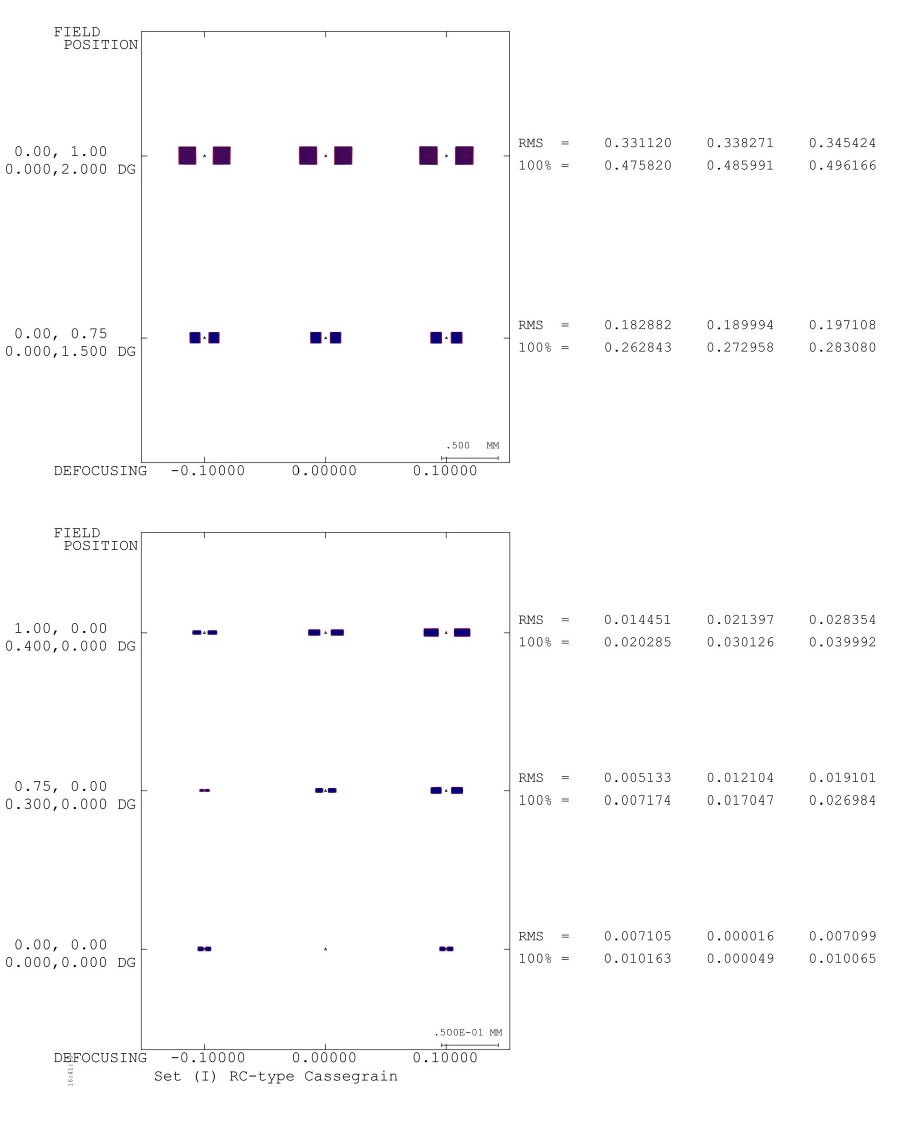}
	\caption
	{The spot diagram for the axially symmetric analog of isolated set (I) with the rectangular aperture 100\,mm\,$\times$\,20\,mm.}
	\label{fig:set1spot_rect}
\end{figure}

\begin{figure}[ht!]
	\centering\includegraphics{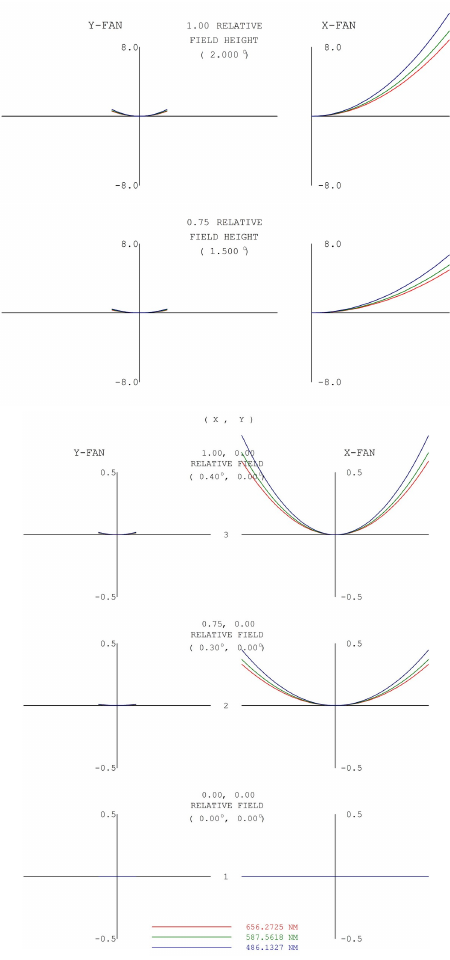}

	\vspace{3mm}
	\caption
	{The OPD plot for the axially symmetric analog of isolated set (I) with the rectangular aperture 100\,mm\,$\times$\,20\,mm.}
	\label{fig:set1OPD_rect}
\end{figure}

Next we analyze performances of the axially symmetric analogs of isolated sets (I) and (II), which perform similarly to the original sets in the XZ and YZ planes, correspondingly. At the same time these modified sets possess focal points that allows for straightforward comparison with the AT performance by means of spot diagram and OPD plot. The perspective view, spot diagram and OPD plot for such analog of set (I) with quadratic $L_x\times L_x$ aperture are shown in Figs.~\ref{fig:set1}, \ref{fig:set1spot} and \ref{fig:set1OPD}, respectively. 
The perspective view, spot diagram and the OPD plot for the analog of set (II) with quadratic $L_y\times L_y$ aperture are shown in Figs.~\ref{fig:set2}, \ref{fig:set2spot} and \ref{fig:set2OPD}, correspondingly. The black circumferences in Fig.~\ref{fig:set2} represent contours of Airy disk, which size is $2.44\times\lambda\times10$ with 10 standing for the focal ratio. Next we track the effect of rectangular pupil shape on performance of the set (I). The perspective view, spot diagram and OPD plot for the axially symmetric analog of set (I) with rectangular $L_x\times L_y$ aperture are shown, respectively, in Figs.~\ref{fig:set1_rect}, \ref{fig:set1spot_rect} and \ref{fig:set1OPD_rect}.
It is clear from the above spot diagrams and OPD plots that for near-axis field points aberrations of the considered AT are mainly originated from its refractive set (II). However, for the object angle of 0.3$^\circ$ and higher the aberrations associated with the reflective set (I) become more essential. 
We pay readers' attention on the different scale for the upper entries in Figs.~\ref{fig:set1spot} and \ref{fig:set1spot_rect}, and notice drastic increase of a spot size with the object angle increase for an isolated the set (I). The introduction of the set (II) allows us to essentially suppress this effect and reduce X component of transverse aberration for the AT, and decisively improve its image quality.

We remark that spot diagram of the anamorphic analog of isolated set (I) in Fig.~\ref{fig:set1spot_rect} possesses two mirror symmetries with respect to the axes X and Y. However, spot diagram of the AT in Fig.~\ref{fig:spot} conserves only one mirror symmetry even for slightly off-axial fields in the XZ plane.


\begin{figure}[ht!]
	\centering\includegraphics{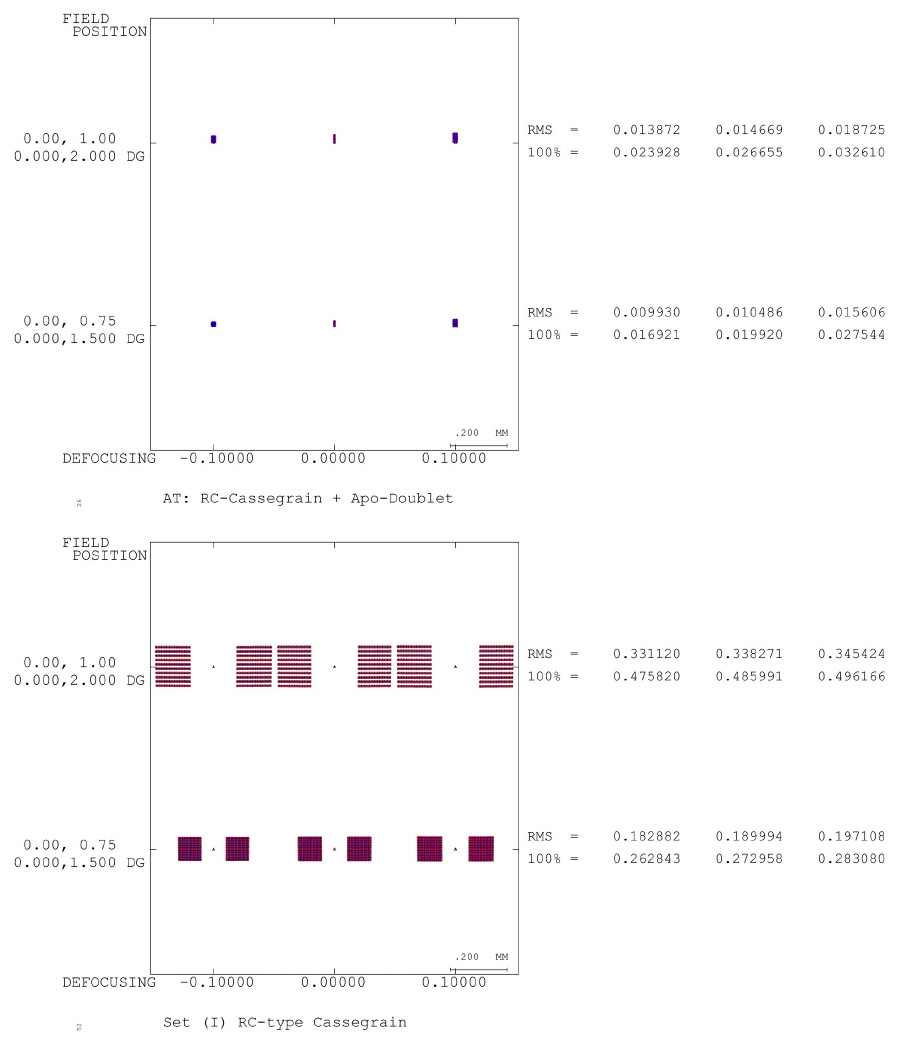}

	\vspace{3mm}
	\caption
	{The spot diagrams for the anamorphic telescope ({\it upper}) and the axially symmetric analog of its isolated set (I) ({\it lower}), for the same rectangular aperture and plotting scale.}
	\label{fig:spot_comparison}
\end{figure}

\subsection{Field of view of the anamorphic telescope}\label{sec:FOV}

Let's compare field of view (FOV) of a usual rotationally symmetric telescope (RST) with the focal distance $F_1$ and FOV of the discussed AT possessing the two focal distances $F_1$ and $F_2$, where $F_1$ is larger than $F_2$. Suppose that they are equipped with the identical eyepieces, which are specified by the focal distance $f$ and 
quadratic field stop, i.e. apparent FOV $\varphi\times\varphi$. 
Then FOV of RST 
is equal angle:
\begin{eqnarray}
	\Phi\times\Phi = \frac{\varphi f}{F_1}\times\frac{\varphi f}{F_1},
\end{eqnarray}
while FOV of AT is $F_1/F_2$ times wider in the direction of the focal segment of the set (I):
\begin{eqnarray}
	\Phi_1\times\Phi_2 = \frac{\varphi f}{F_1}\times\frac{\varphi f}{F_2}.
\end{eqnarray}
We conclude that AT possesses $F_1/F_2$ times wider FOV in the Y direction with respect to RST of the same rectangular aperture. 
In particular, in the above examples of the AT (Figs.~\ref{fig:simulation} and \ref{fig:simulation_per}) and the RST (Fig.~\ref{fig:set1_rect}), which are identical in size, the former has five times wider FOV in the Y direction with respect to the later if both are equipped with identical eyepieces. 
We remark that in case of RST essential increase of FOV in the Y direction results in dramatic growth of aberrations that is shown in Fig.~\ref{fig:spot_comparison}, which compares the spot diagrams for AT and RST for the same two fields with nonzero Y object angles. 
Similarly, in case of using a rectangular CCD camera or 
active-pixel sensor, AT provides $F_1/F_2$ times wider FOV in the Y direction, along which magnification is $F_1/F_2$ times lower.

\section{Discussion and conclusion}\label{sec:4}

After a brief intro on anamorphic OS aberrations and their 
partial compensation, we 
presented the following application. 
We proposed an anamorphic catadioptric telescope that combines 
translationally invariant reflective part and an anamorphic refractive part, which has a rotational symmetry about the 
focal segment of the reflective part. 
We outlined derivation of analytical expressions for the GS aberration compensations 
in the OS symmetry planes, in the case of 
cylindrical version of the Cassegrain reflector as the reflective part and two 
alternatives for the refractive part, which are relatively simple for analytical consideration. Then we simulated the anamorphic 
telescope combining the cylindrical RC-type Cassegrain with toroidal apochromatic doublet, using the ray-tracing 
software CODE~V~\cite{CODE:V}. This simulation 
demonstrated good OS performance, which can be further improved by using more advanced refractive set and introducing corrective lenses. 

In general, a wide class of cylindrical versions of known telescopic objectives, reflective, refractive or catadioptric, can serve as a set (I) and variety of generalized toroidal versions of known axially symmetric reflective, refractive or catadioptric telescopic OS can serve as a set (II) 
(DZ, patent submission Nr.\,P.452264 to the Patent Office of the Republic of Poland, 2025). 
Moreover, the presented method of separate focusing in the two dimensions can be extended to off-axis systems.
 
One of possible limitations for the introduced anamorphic OS comes from avoiding light blocking by the components of set (II), e.g., the distance to the focus behind the secondary mirror must be sufficiently greater than the focal distance $F_2$ of the set (II). However, this issue is not crucial for a wide class of solutions, in particular, such blocking does not take place for the examples shown in section~\ref{sec:proof}. 
The expressions for aberration compensation for the OS symmetry planes XZ and YZ (in the discussed coordinates) are identical to the ones for axially symmetric analogs of the sets (I) and (II), respectively. Aberration compensation beyond these planes, including chromatic and field curvature aberrations, is an important direction of future research. Several simplifications leave us with a simple two-step design procedure for the refractive set (II): (a) all parameters for the YZ plane (refractive indices, radii of curvature, lens thicknesses and separations, etc.) can be derived using the same methods as for ordinary rotationally symmetric OS, (b) each optical element of the set (II) can be obtained as the body of rotation of a given YZ profile with respect to the focal line of the set (I) by a proper angle to collect the converging light beam from the set (I).

The introduced optical scheme can be utilized in the design of terrestrial and space telescopes of various functional purposes, which would require inclusion of FOV and optimization to correct for coma, etc. However, the goal of this paper is a concept presentation of a new OS class rather than a complete design of its particular realization.

Obvious benefits of the discussed anamorphic telescope, such as weight saving and fittability, arise from the shape of its primary mirror. 
A rectangular shape of a primary in the XY projection can provide economy in mass for the same resolution in the X (larger side) direction compared to a circular shape, while rotation of the telescope along the Z axis governs direction of its best resolution. 
The size of mass saving can be easily estimated. Assuming that mass density and thickness of the primaries of a rotationally symmetric telescope and an anamorphic one are the same and constant, the ratio of masses of these mirrors, which essentially contribute to the masses of the instruments, is equal to the ratio of their surface areas, $M_{\rm RST}/M_{\rm AT}=S_{\rm RST}/S_{\rm AT}$. 
Taking the diameter of the RST's circular aperture to be equal to the larger side $L_x$ of the AT's rectangular aperture $L_x\times L_y$, for the case of circular $XZ$ profiles the ratio of their surface areas can be written as 
\begin{flalign}
	\frac{S_{\rm RST}}{S_{\rm AT}} = \frac{\pi h}{L_y\arccos\left( \frac{R-h}{R} \right)},
\end{flalign}
where $h$ is the depth (sag) and $R$ is the radius of curvature 
in the XZ plane. (Strictly speaking, one has to compare instruments of equal maximal resolution rather then equal maximal aperture size~\cite{Zhuridov:patent}. Modification of the Rayleigh criterion for the anamorphic OS, which we ignore here, may further increase the discussed gain in mass.) Using the relation
\begin{flalign}
	L_x = 2\sqrt{2Rh-h^2},
\end{flalign}
for small $h$ with respect to $R$ we obtain an estimate for the ratio of masses of the primaries in the the following simple form
\begin{flalign}
	\frac{M_{\rm RST}}{M_{\rm AT}} \approx  \frac{\pi}{4} \frac{L_x}{L_y},
\end{flalign}
which is equal to the ratio of respective aperture areas as expected. A more narrow AT's primary provides larger gain in mass. However, a more 
distorted image is generated that complicates reconstruction of the object's appearance.

Potentially, the discussed two-step focusing scheme can be useful also when designing solar observatories. 
Additionally, reduction of manufacturing costs is possible for a cylindrical primary (in particular, produced by bending a flat mirror surface) compared to a rotationally symmetric one. However, adaptation of polishing technology and testing methods is required. 

One of more traditional approaches to weight saving without loss of aperture size is in taking a narrow slice of a rotationally symmetric primary. This method can significantly reduce costs in case of large but stationary aperture. An interesting solution is RATAN-600 telescope, which primary comprises a 576 m diameter circle of rectangular radio reflectors that can be angled causing the overall effect of a partially steerable antenna. This design works well for the radio waves. However, it is hard to apply it for much shorter wavelengths that would require much higher accuracy in positioning and shaping the reflecting parts. Concerning possible space applications, this approach can benefit in weight saving, but would suffer from dimensional restrictions existing for payloads. On the contrary, the introduced anamorphic solution allows for fully steerable ground-based telescopes with large aperture in one dimension. Moreover, in space applications, it reconciles large aperture (maximal) size, large enough effective aperture area and good fittability to either certain satellite formats or rocket cargo bays. 
Indeed, a telescope of this type can be made nearly flat with relatively small Y side. Then it can be shortened 
in the Z direction using one or more tilting mirrors, in particular, a flat mirror located behind the primary, which 
redirects the beam along the primary's longer side. In result, the larger sides of the telescope, $A_x$ and $A_z$, are essentially determined by the primary's length $L_x$ and distance between the primary and secondary mirrors, correspondingly. In particular, such OS can be fitted within the 6U, 9U and other specific small satellite formats, where 1U 
is a standard CubeSat size measuring $10\times10\times10$~cm.

To conclude, we presented a generic concept of an anamorphic telescope of a new type that contains cylindrical and toroidal optical surfaces, and can possess two large magnifications in the orthogonal directions of the image plane corresponding to the two independent focal distances, which allow to control both dimensions of the filed of view. The primary of this telescope can be contracted in one dimension keeping reasonably small aberrations and avoiding reduction of its field of view. This concept is promising for the sector of space optical industry, where the instrument’s shape and mass are of particular importance.

\subsection*{Disclosures}
The author declares no conflicts of interest.
\subsection* {Code, Data, and Materials Availability} 
All data in support of the findings of this paper are available within the article.
\subsection* {Acknowledgments}
The author thanks Barbara Pintaske and Daniela Ponce for their attention when providing the optical software of the SYNOPSYS, and the JATIS Reviewers for their useful comments.
%


\bibliography{paper2refs}   
\bibliographystyle{spiejour}   

%


\end{spacing}
\end{document}